\documentclass[aps,prd,twocolumn,preprintnumbers,%superscriptaddress,
floatfix,nofootinbib]{revtex4}

\usepackage{graphicx,amssymb,amstext,mathrsfs,dsfont,amsfonts}
\usepackage{eucal,wrapfig,boxedminipage,setspace,subfigure,epsfig}
\usepackage{color,latexsym,url}
\usepackage[colorlinks,hyperindex]{hyperref}
    \hypersetup
    {
        colorlinks,%
        citecolor=red,%
        linkcolor=blue,%
        urlcolor=black,%
    }

\def\a  {\alpha}                
       \def\d  {\delta}        
\def\e  {\epsilon}

 \newcommand{\call}{\mbox{${\cal L}$}}

\def\IR{{\hbox{{\rm I}\kern-.2em\hbox{\rm R}}}}
\def\IB{{\hbox{{\rm I}\kern-.2em\hbox{\rm B}}}}
\def\IN{{\hbox{{\rm I}\kern-.2em\hbox{\rm N}}}}
\def\IC{\,\,{\hbox{{\rm I}\kern-.59em\hbox{\bf C}}}}
\def\IZ{{\hbox{{\rm Z}\kern-.4em\hbox{\rm Z}}}}
\def\IP{{\hbox{{\rm I}\kern-.2em\hbox{\rm P}}}}
\def\IH{{\hbox{{\rm I}\kern-.4em\hbox{\rm H}}}}
\def\ID{{\hbox{{\rm I}\kern-.2em\hbox{\rm D}}}}

\def\be{\begin{equation}}
\def\ee{\end{equation}}
\def\ba{\begin{eqnarray}}
\def\ea{\end{eqnarray}}

\def\half{\frac{1}{2}}

\def\ra{\rightarrow}

\def\del{\partial}

\newcommand{\brac}[1]{\langle #1 \rangle}

\def\nn{\nonumber}
\def\ea{{\it et al}. }

\newcommand{\wt}{\widetilde}

\newcommand{\beq}{\begin{equation}}
\newcommand{\eeq}{\end{equation}}
\newcommand{\bea}{\begin{eqnarray}}
\newcommand{\eea}{\end{eqnarray}}

\newcommand{\tw}{{{\widetilde w}}}

\newcommand{\trho}{{{\widetilde \rho}}}
\newcommand{\td}{{{\widetilde d}}}
\newcommand{\tL}{{{\widetilde L}}}
\newcommand{\tc}{{{\widetilde c}}}
\newcommand{\tm}{{{\widetilde m}}}
\newcommand{\tK}{{{\widetilde K}}}
\newcommand{\tmu}{{{\widetilde \mu}}}
\newcommand{\tF}{{{\widetilde F}}}
\newcommand{\tOmega}{{{\widetilde \Omega}}}
\newcommand{\tE}{{{\widetilde E}}}

\begin{document}

%\begin{frontmatter}

%\voffset 2.5cm

\newcommand\sect[1]{\emph{#1}---}

%\pagestyle{empty}

%\preprint{
%\begin{minipage}[t]{3in}
%\begin{flushright} SHEP-10-01
%\\[30pt]
%\hphantom{.}
%\end{flushright}
%\end{minipage}
%}

\title{Phase diagram of the D3/D5 system in a magnetic field and a BKT transition}

\author{Nick Evans}
\email{evans@soton.ac.uk}
\author{Astrid Gebauer}
\email{ag806@soton.ac.uk}
\author{Keun-Young Kim}
\email{k.kim@soton.ac.uk}
\author{Maria Magou}
\email{mm21g08@soton.ac.uk}

\affiliation{ School of Physics and Astronomy, University of
Southampton, Southampton, SO17 1BJ, UK \\ % \vspace{0.2cm}
}

\begin{abstract}
\noindent We study the full temperature and chemical potential
dependence of the D3/D5 2+1 dimensional theory in the presence of
a magnetic field. The theory displays separate transitions
associated with chiral symmetry breaking and melting of the bound
states. We display the phase diagram which has areas with first
and second order transitions meeting at two critical points
similar to that of the D3/D7 system. In addition there is the
recently reported BKT transition at zero temperature leading to
distinct structure at low temperatures.
\end{abstract}
%\date{\today}

%\begin{keyword}


%\end{keyword}

%\end{frontmatter}
\maketitle

%\tableofcontents

%
\begin{figure*}[]
\centering
 \subfigure[{\small $\,$ D3/D7 }]
  {\includegraphics[width=7cm]{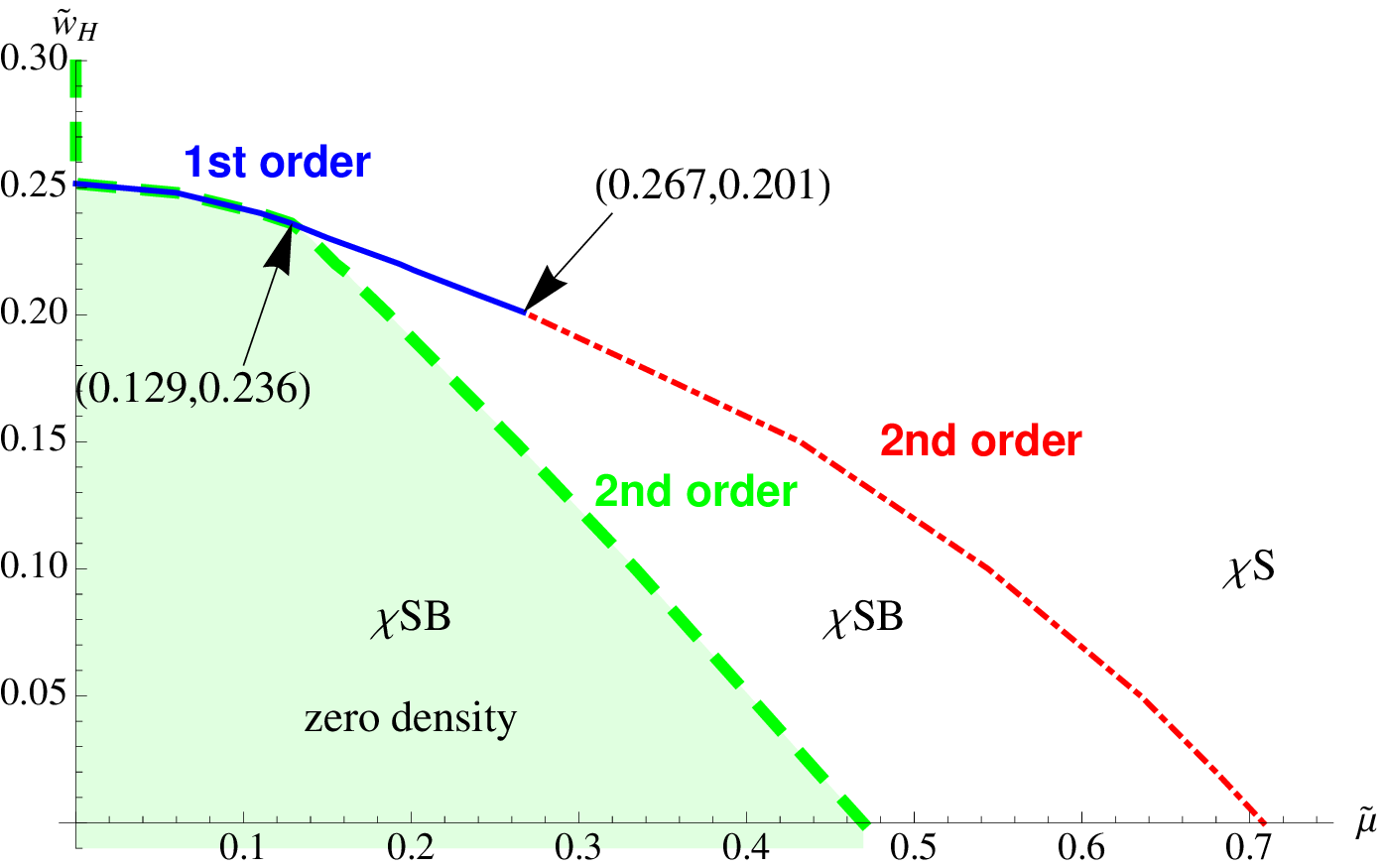}} \ \ \ \ \ \ \ \ \ \
  \subfigure[{\small $\,$ D3/D5}]
  {\includegraphics[width=7cm]{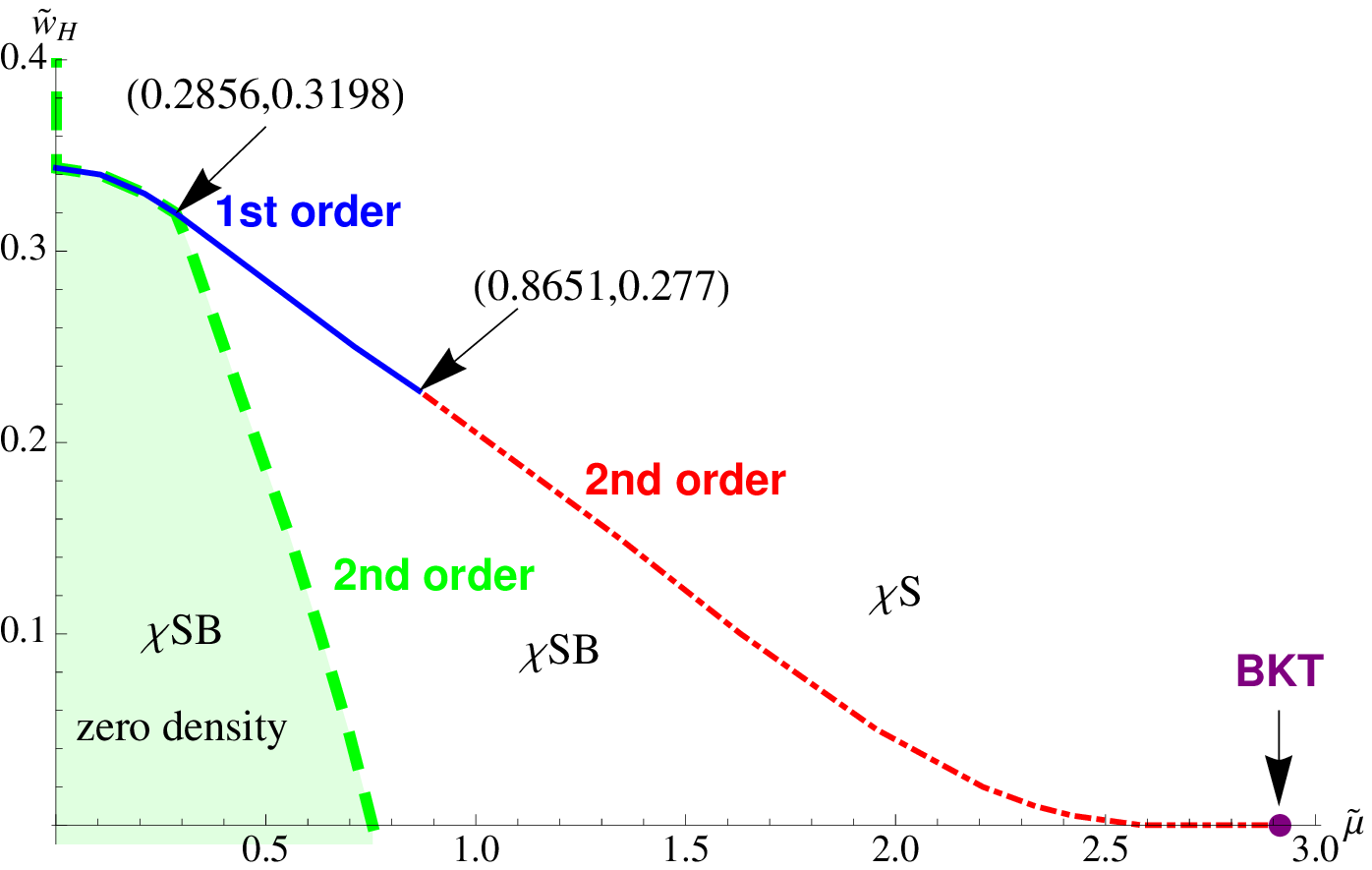}} \ \ \ \ \ \ \ \ \ \
  \caption{
           {\small  The phase diagrams for the D3/D7~\cite{Evans1} and D3/D5 systems. $\tilde{w}_H$
           measure the temperature of the theory whilst $\tilde{\mu}$ is the chemical potential.
           The dashed line is a second order transition associated with the formation
           of quark density and meson melting. The dotted line is a second order transition
           for chiral symmetry restoration. In the D3/D5 case that transition ends at a BKT
           transition point and its effects on the second order line can be seen. The continuous
           line is the merged first order transition.  The position of critical points are marked.   }
           }\label{Fig1}
\end{figure*}

\section{introduction}

There has been recent interest in holographic descriptions of the
phase structure of gauge theories in the presence of magnetic
fields~\cite{Johnson1,Albash:2007bk,Magnetic,Evans1,Karch1,Filev1,Karch2,Liu}.
The D3/D7 holographic system describes a confining 3+1d gauge
theory with quarks~\cite{Rev}. The magnetic field induces chiral
symmetry breaking. The symmetry breaking and quark confinement are
lost at high temperature and density. Between is a rich structure
of phase transitions of both first and second order meeting at
critical points. These transitions have been explored
in~\cite{Evans1} and the summary phase diagram is displayed in Fig
\ref{Fig1}a. Here the theory is interesting as a loose analogue
for QCD which is also a confining and chiral symmetry breaking
gauge theory but where we can not as yet compute the precise phase
diagram.

Interest has also turned to the D3/D5 system \cite{d3d5} that
describes fundamental representation matter fields on a 2+1d
defect within a 3+1d gauge theory. This system may have some
lessons for condensed matter systems. In~\cite{Karch2} an analysis
of the D3/D5 system at finite density ($d$) and at zero
temperature ($T$) revealed that the chiral symmetry breaking
transition with increasing magnetic field ($B$) is not second
order but similar to a Berezinskii-Kosterlitz-Thouless (BKT)
transition~\cite{BKT} (see also the holographic example in
\cite{Kaplan,Liu}). That is order parameters across the transition grow
as ${\rm exp} (-a/\sqrt{\nu_c-\nu})$ where $a$ is a constant and
$\nu = d/B$. ($\nu_c$ the critical value for the transition). For
small T the authors of~\cite{Karch2} showed the BKT transition
returns to a second order nature. This difference from the D3/D7
case is surprising so it seems worth fleshing out the entire phase
diagram for the theory to see if other surprises are present. In
this letter we present that analysis - much of the computation
matches that in the D3/D7 system which we worked through in detail
in~\cite{Evans1} so here we very briefly present the formalism and
the conclusions. We display the resulting phase diagram for
massless matter fields in Fig \ref{Fig1}b. Clearly much of the
structure is similar to the D3/D7 case but the second order
boundary of the chiral symmetry breaking phase is distorted by the
presence of the BKT transition.

\section{The holographic description}

The ${\cal N}$=4 super Yang-Mills gauge theory at finite
temperature has a holographic description in terms of an AdS$_5$
black hole geometry (with $N$ D3 branes at its
core)\cite{Malda}. The geometry can be
written as
\begin{eqnarray}
   ds^2 &=& \frac{w^2}{R^2}(- g_t dt^2 + g_x d\vec{x}^2 + g_x dy^2) \nn \\
        & & + \frac{R^2}{w^2} (d\rho^2 + \rho^2 d\Omega_2^2
         + dL^2 + L^2 d\bar{\Omega}_2^2) \ ,
\end{eqnarray}
where $\vec{x}$ is two dimensional, $y$ will be the D3 coordinate
not shared by our D5, we have split the transverse six plane into
two three planes each with a radial coordinate $\rho,L$ and a two
sphere, $R^4=4 \pi g_s N \alpha^{'2}$ and
\begin{eqnarray}
g_t := \frac{(w^4 - w_H^4)^2}{2 w^4 (w^4+w_H^4)}\ ,  \qquad g_x :=
\frac{w^4 + w_H^4}{ 2 w^4} \ .
\end{eqnarray}
The temperature of the theory is given by the position of the
horizon, $w_H = \pi R^2 T$

%Quenched ($N_f \ll N$) ${\cal N}$=2 quark superfields can be
%included in the ${\cal N}$=4 gauge theory through probe D7 branes
%in the geometry\cite{Karch,Polchinski,Bertolini:2001qa,Mateos}.
%The D3-D7 strings are the quarks. D7-D7 strings holographically
%describe mesonic operators and their sources.
We include our 2+1d defect with fundamental matter fields by
placing a probe D5 brane in the D3 geometry. The probe limit
corresponds to the quenched limit of the gauge theory. The D5
probe can be described by its DBI action
\beq S_{DBI} = - T_{D5} \int d^6\xi \sqrt{- {\rm det} (P[G]_{ab} +
2 \pi \alpha' F_{ab})} \ , \eeq
where $P[G]_{ab}$ is the pullback of the metric and $F_{ab}$ is
the gauge field living on the D5 world volume. We will use
$F_{ab}$ to introduce a constant magnetic field (eg $F_{12} = -
F_{21} = B$)~\cite{Johnson1} and a chemical potential associated with baryon
number $A_t(\rho) \neq 0$~\cite{Myers1,Kim}
We embed the D5 brane in the $t,\vec{x},\rho$ and $ \Omega_2$ directions of
the metric but to allow all possible embeddings must include a
profile $L(\rho)$ at constant $y,\bar{\Omega}_2$. The full DBI action we
will consider is then
\begin{eqnarray}
  S = \int d\xi^6 \call(\rho)
    = \left(\int_{S^2} \e_2 \int dtd\vec{x} \right) \int d\rho \
  \call(\rho) \ ,
\end{eqnarray}
where $\e_2$ is a volume element on the 2-sphere and
\begin{eqnarray}
  \call &:=& -N_f T_{D5} \frac{\rho^2}{2\sqrt{2}}\left(1-\frac{w_H^4}{w^4}\right)
  \nn \\
  &&\times \sqrt{\left(1+(\partial_\rho L)^2
   - \frac{ 2 w^4 (w^4+w_H^4)}{(w^4 - w_H^4)^2} (2\pi\a' \partial_\rho A_t)^2 \right)}
   \nn \\
  &&\times \sqrt{\left(\left(1+\frac{w_H^4}{w^4}\right) + \frac{4 R^4}{w^4+w_H^4}B^2 \right)} \ . \label{OriginalAction}
\end{eqnarray}
Since the action is independent of $A_t$, there is a conserved
quantity $d$ $\left(:= \frac{\delta S}{\delta F_{\rho t}}\right)$
and we can use the Legendre transformed action
\begin{eqnarray}
  \!\!\!\! \!\!\!\! \wt{S} = S - \int d\xi^6 F_{\rho t} \frac{\delta S}{\delta F_{\rho t}}
         =  \left(\int_{S^2} \e_2 \int dtd\vec{x} \right) \int d\rho \
  \wt{\call}(\rho) \ ,  \label{LegendreAction}
\end{eqnarray}
where
\begin{eqnarray}
  &&\wt{\call} := - N_f T_{D5} \frac{(w^4-w_H^4)}{2\sqrt{2} w^4}
  \sqrt{K (1+(\partial_\rho L)^2)} \label{Hamiltonian}  \\
      && K  :=  \left(\frac{w^4+w_H^4}{w^4}\right) \rho^4
    + \frac{4  R^4 B^2}{w^4+w_H^4} \rho^4 \nn \\
    && \qquad + \frac{4 w^{4}}{(w^4+w_H^4)}
    \frac{d^2}{(N_f T_{D5} 2\pi\a')^2} \ .
\end{eqnarray}
To simplify the analysis we note that we can use the magnetic
field value as the intrinsic scale of conformal symmetry breaking
in the theory - that is we can rescale all quantities in
(\ref{Hamiltonian})  by $B$ to give
\begin{eqnarray} \label{rescale}
  &&\!\!\!\!\!\!\!\!\!  \wt \call = -N_f T_{D3} (R\sqrt{B})^3 \
        \frac{\tw^4 - \tw_H^4}{\tw^4}\sqrt{\tK(1+\tL'^2)}\ ,
         \\
   &&\!\!\!\!\!\!\!\!\!  \tK = \left(\frac{\tw^4+\tw_H^4}{\tw^4}\right) \trho^4
                +\frac{1}{\tw^4+\tw_H^4} \trho^4
              + \frac{\tw^4}{(\tw^4+\tw^4_H)} \td^2 \ , \label{tH}
\end{eqnarray}
where the dimensionless variables are defined as
\begin{eqnarray}
   && (\tw , \tL, \trho, \td)  \\
   && := \left(\frac{w}{R\sqrt{2B}},
        \frac{L}{R\sqrt{2B}} ,
       \frac{\rho}{R\sqrt{2B}},
       \frac{d}{(R \sqrt{B})^2 N_f T_{D5} 2\pi \a'}\right) \ . \nn
\end{eqnarray}

In all cases the embeddings become flat at large $\rho$ taking the
form
\begin{eqnarray}
   \tL(\trho) \sim \tm + \frac{\tc}{\trho}\ ,\label{scaled}
  % \tm = \frac{2\pi \a' m_q}{R\sqrt{2B}} \ ,
  % \tc = \brac{\bar{q}q} \frac{(2\pi \a')^3}{(R\sqrt{2B})^3} \ . \
\end{eqnarray}
In the absence of temperature, magnetic field and density the
regular embeddings are simply $L(\trho)= \tm$, which is the
minimum length of a D3-D5 string, allowing us to identify it with
the quark mass as shown. $\tilde{c}$ should then be identified
with the quark condensate.

We will classify the D5 brane embeddings  by their small $\trho$
behavior. If the D5 brane touches the black hole horizon, we call
it a black hole embedding, otherwise, we call it a Minkowski
embedding.  We have used Mathematica to solve the equations of
motion for the D5 embeddings resulting from (\ref{rescale}).
Typically in what follows, we numerically shoot out from the black
hole horizon (for black hole embeddings) or the $\trho=0$ axis
(for Minkowski embeddings) with Neumann boundary condition for a
given $\td$. Then by fitting the embedding function with
(\ref{scaled}) at large $\trho$ we can read off $\tm$ and $\tc$.

The Hamilton's equations from (\ref{LegendreAction}) are
$\del_\rho d = \frac{\d\wt{S}}{\d A_t} $ and $\del_\rho A_t = -
\frac{\d\wt{S}}{\d d} $. The first simply means that $d$ is the
conserved quantity. The second reads as
\begin{eqnarray} \label{muder}
  \del_\trho \wt{A}_t = \td\ \frac{\tw^4 - \tw_H^4}{\tw^4 + \tw_H^4} \sqrt{\frac{ 1+(\tL')^2 }{\tK}} \ ,
\end{eqnarray}
where $\wt{A}_t := \frac{\sqrt{2}2\pi\a'A_t}{R\sqrt{2B}}$.

There is a trivial solution of (\ref{muder}) with $\td =0$ and
constant $\wt{A}_t$~\cite{Myers3}. The embeddings are then the
same as those at zero chemical potential. For a finite $\td$,
$\tilde{A}_t'$ is singular at $\trho=0$ and requires a source. In
other words the electric displacement must end on a charge source.
The source is the end point of strings stretching between the D5
brane and the black hole horizon. The string tension pulls the D5
branes to the horizon resulting in black hole
embeddings~\cite{Myers1}.  For such an embedding the chemical
potential($\tmu$) is defined as
\begin{eqnarray}
  \tmu &:=&\lim_{\trho \ra \infty} \wt{A}_t(\trho) \nn \\
  &=& \int_{\trho_H}^\infty d\trho\ \td\ \frac{\tw^4 - \tw_H^4}
  {\tw^4 + \tw_H^4} \frac{\sqrt{ 1+(\tL')^2} }{\sqrt{\tK}} \ , \label{mu}
\end{eqnarray}
where we fixed $\wt{A}_t(\trho_H) = 0$ for a well defined $A_t$
at the black hole horizon.

The generic analysis below with massless quarks and $B$, $T$ and
$\mu$ all switched on involve four types of solution of the Euler
Lagrange equations. All of these approach the $\tilde{\rho}$ axis
at large $\rho$ to give a zero quark mass. Firstly, there are
Minowski embeddings that avoid the black hole so have a non-zero
condensate $\tc$ - these solutions have $\tilde{d}=0$ so
$\tilde{A}_t=\mu$. Secondly, there can be generic black hole
solutions with both of $\tc$ and $\td$ none zero. Finally there
are solutions that lie entirely along the $\tilde{\rho}$ axis so
that $\tc=0$ but with $\td$ either zero or non zero. In fact the
flat embeddings with $\td=0$ are always the energetically least
preferred but the other three all play a part in the phase diagram
of the theory.

To compare these solutions we compute the relevant thermodynamic
potentials. The Euclideanized on shell bulk action can be
interpreted as the thermodynamic potential of the boundary field
theory. The Grand potential ($\tOmega$) is associated with the
action (\ref{OriginalAction}) while the Helmholtz free energy
($\tF$) is associated with the Legendre transformed action
(\ref{LegendreAction}):
\begin{eqnarray}
  && \tF(\tw_H,\td) :=   \frac{-\wt{S}}{N_f T_{D5} (R\sqrt{B})^3
  \mathrm{Vol}} \nn \\
  && \quad = \int_{\trho_H}^\infty d\trho\  \frac{\tw^4 - \tw_H^4}{\tw^4 }
  \sqrt{(1+(\tL')^2)}\sqrt{\tK} \label{F} \\
  && \tOmega(\tw_H,\tmu) :=  \frac{-{S}}{N_f T_{D5} (R\sqrt{B})^3
   \mathrm{Vol}} \nn \\
  && \quad = \int_{\trho_H}^\infty d\trho\  \frac{\tw^4 - \tw_H^4}{\tw^4 }
  \sqrt{(1+(\tL')^2)} \frac{\tK(\td=0)}{\sqrt{\tK}}  \label{Omega}
\end{eqnarray}
where $\mathrm{Vol}$ denote the trivial 5-dimensional volume integral
except $\trho$ space, so the thermodynamic potentials defined
above are densities, strictly speaking. Since $\tK \sim \trho^4$,
both integrals diverge as $\trho^2$ at infinity and need to be
renormalized.

%The same form as the thermodynamic potentials, (\ref{mu}),(\ref{F})
%and (\ref{Omega}) hold also for D3/D7 case with a different $\tK$,
%
%\begin{eqnarray}
%  && D5:\ \left(\frac{\tw^4+\tw_H^4}{\tw^4}\right) \trho^4
%                +\frac{1}{\tw^4+\tw_H^4} \trho^4
%              + \frac{\tw^4}{(\tw^4+\tw^4_H)} \td^2 \ , \nn \\
%  && D7:\ \left(\frac{\tw^4+\tw_H^4}{\tw^4}\right)^2 \trho^6
%                +\frac{1}{\tw^4} \trho^6
%              + \frac{\tw^4}{(\tw^4+\tw^4_H)} \td^2 \ .
%\end{eqnarray}

%
\begin{figure}[]
\centering
  \includegraphics[width=7cm]{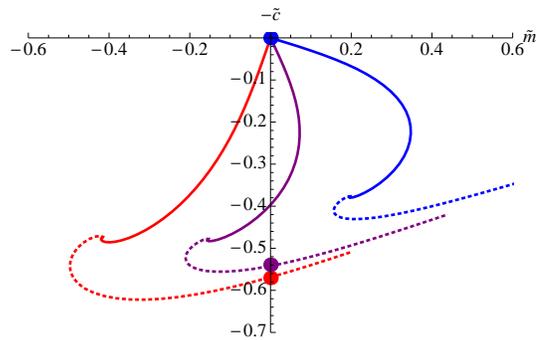}
  \caption{
           {\small A plot of the condensate vs the quark mass to show
           the first order phase transition at zero chemical potential induced by temperature.
           The solid line corresponds to the black hole embedding and the dotted  line to
           a Minkowski embedding.
           From bottom to top the curves correspond to temperatures $\tw_H = 0.25, 0.3435, 0.45$.
           }} \label{Fig2}
\end{figure}

\section{Chiral symmetry restoration by temperature}

The chiral symmetry restoration transition by temperature is first
order \cite{Filev1} (a transition related to the thermal
transition for non-zero mass at B=0 ~\cite{Babington}). The
transition on the gravity side is between a Minkowski embedding
that avoids the black hole to an embedding that lies along the
$\tilde{\rho}$ axis ending on the black hole. Fig \ref{Fig2} shows
the $(-\tc,\tm)$ diagram for some temperatures ($\tw_H = 0.25,
0.3435, 0.45$ from the bottom). The solid lines are the black hole
embeddings and the dotted lines are Minkowski embeddings. Since we
are interested in the case $\tm=0$, the condensate is the
intersect of the curves with the vertical axis. As temperature
goes up the condensate moves from the lower dot to the middle
curve continuously, then jumps at $\tw_H =0.3435$ to the origin
(zero condensate), which corresponds to the chiral symmetric
phase. It is also the transition from a Minkowski (dotted line) to
a black hole embedding (solid line). This jump can be seen by a
Maxwell construction: $\tm$ and $\tc$ are conjugate variables and
the two areas between the middle curve and the axis are equal at
the transition point. See \cite{Filev1} for more details.

This transition as well as restoring chiral symmetry also
corresponds to the melting of bound states of the defect quarks
since the Minkowski embedding has stable linearized mesonic
fluctuation whilst the black hole embedding has a quasi-normal
mode spectrum~\cite{Melting}.

\begin{figure*}[]
\centering
  \subfigure[{\small $\,$ $\tw_H = 0.15 $ }]
  {\includegraphics[width=5cm]{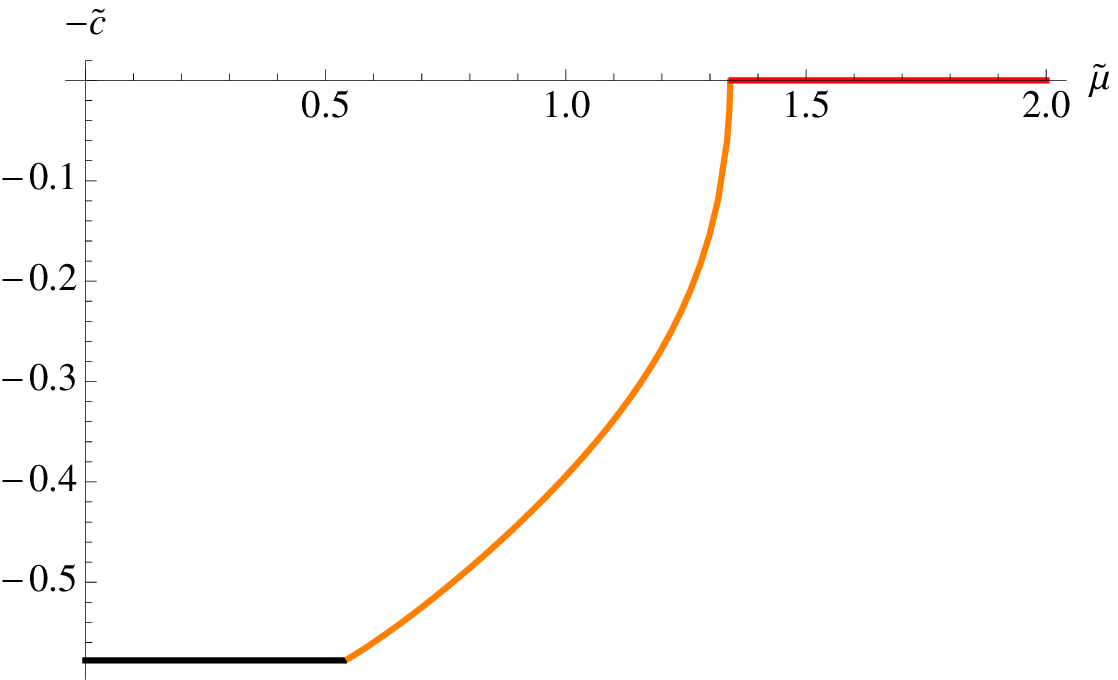}} \ \ \ \
  \subfigure[{\small $\,$ $\tw_H =  10^{-5} $ }]
  {\includegraphics[width=5cm]{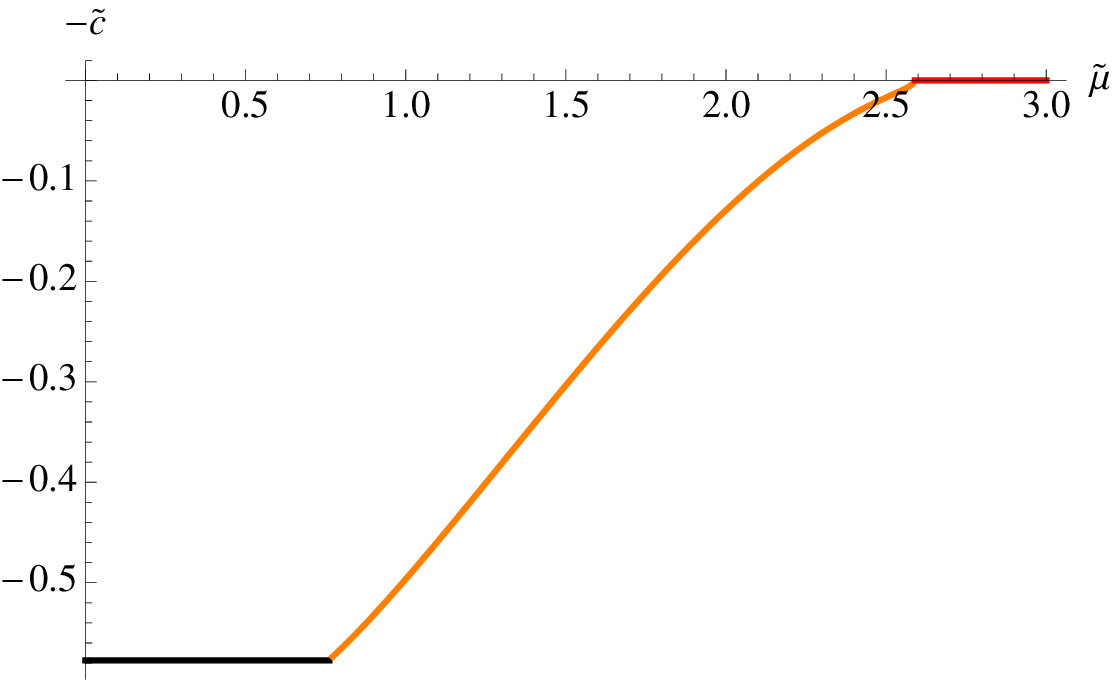}}
  \subfigure[{\small $\,$ $\tw_H = 10^{-5}$ near the critical point }]
  {\includegraphics[width=5cm]{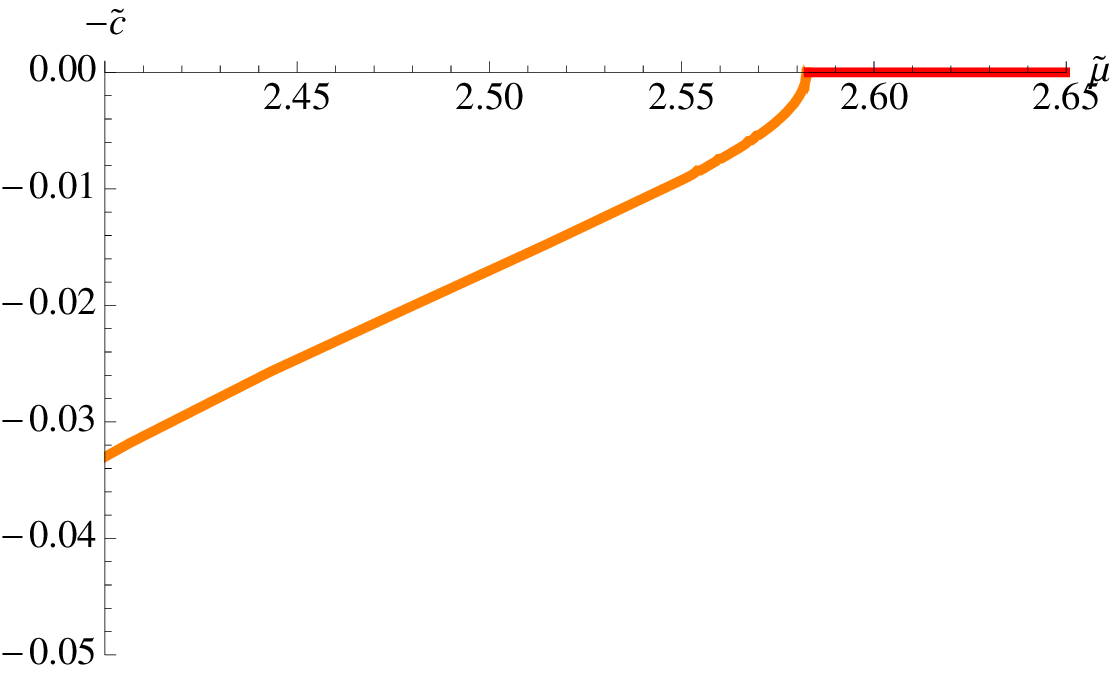}} \ \ \ \
  \caption{
           {\small Plots of the condensate vs chemical potential on fixed temperature
           slices, showing the phase structure of the theory.
           Figure (b) and (c) show that at low temperature the BKT
           transition becomes second order.
            }
           }\label{Fig3}
\end{figure*}

\section{Chiral symmetry restoration by density}

At zero temperature we find two phase transitions with increasing
chemical potential. 

At low chemical potentials the preferred embedding is a Minkowski
embedding with $\tilde{A}_t=\mu$ so there is no quark density.
There is then a transition to a black hole embedding with non-zero
quark density, $\td$. This transition, whilst appearing first
order in terms of the brane embeddings, displays second order
behaviour in all field theory quantities such as the condensate or
density (which grows smoothly from zero). The transition also
corresponds to the on set of bound state melting since the black
hole embedding has quasi-normal modes rather than stable
fluctuations.

The chiral symmetry transition induced by density at zero
temperature is distinct and also a continuous transition. It has
been shown to be of the BKT type for this D3/D5 case~\cite{Karch2}
as opposed to a mean-field type second order transition as seen in
the D3/D7 case \cite{Evans1,Karch1}.

The chiral symmetric phase corresponds to the trivial embedding,
$L=0$. Chiral symmetry breaking is signaled by the instability of
small fluctuation around the $L=0$ embedding. The Free energy
(\ref{F}) with (\ref{rescale}) at zero T reads
\begin{eqnarray}
 \tF &\sim& \sqrt{1+\tL'^2}\sqrt{
 \trho^4 + \frac{\rho^4}{\tw^4}+ \td^2} \ ,
\end{eqnarray}
which can be expanded up to the quadratic order in $\tL$ as
\begin{eqnarray}
 \tF &\sim& -\half \sqrt{1+\trho^4 + \td^2} \tL'^2
 + \frac{\tL^2}{\trho^2\sqrt{1+\trho^4 + \td^2}}
\end{eqnarray}
At $\trho \gg 1$, $\frac{\tL}{\trho}$ behaves as a scalar with
$m^2 = -2$ in AdS$_4$, while at small $\trho \ll 1$ and $\trho \ll
\td$ it behaves as a scalar with $m^2 = -\frac{2}{1+\td^2}$ in
AdS$_2$. The Breitenlohner-Freedman (BF) bound of
AdS$_2$ is $-\frac{1}{4}$, so below  $\td_c = \sqrt{7}$ the BF
bound is violated and the embedding $\tL=0$ is
unstable~\cite{Karch2}. This critical density corresponds to the
critical chemical potential $\tmu \sim 2.9$ as can be computed
from (\ref{mu}). In \cite{Karch2} it was shown that the condensate
scales near this transition as
\begin{eqnarray}
  -\tc \sim -  e^{-  \pi \sqrt{\frac{1+\td^2}{\td_c^2-\td^2}}} \ , \label{BKTc}
\end{eqnarray}
which corresponds to BKT scaling~\cite{BKT}.  This transition is an
example of the analysis in \cite{Kaplan} where it was shown that
if a scalar mass in a holographic model could be tuned through the
BF bound a BKT transition would be seen at the critical point.

\section{Phase diagram in $\mu$-T plane}

To compute the full phase diagram we work on a series of constant
T slices. We have found the four relevant embeddings discussed
above and found those that minimize the relevant thermodynamic
potential. For more details of the method and relevant analysis we
refer to \cite{Evans1}, where we studied D3/D7 system using the
same methods.   Fig 3 shows some example plots of the dependence
of the condensate on the density on fixed T slices. 
%Fig \ref{Fig3}a for example 
It shows that the Minkowski embedding with
$\td=0$ is preferred at low $\tilde{\mu}$, a black hole embedding
with growing $\td$ at intermediate $\tilde{\mu}$, before finally a
transition to a flat embedding occurs at high chemical potential.

Qualitatively the phase diagram, shown in Fig \ref{Fig1}, is almost the
same as the D3/D7 case - the two second order transitions at zero
temperature converge at two critical points to form the first
order transition identified at zero density. The only difference
is induced by the chiral phase transition at zero T. Comparing to
the D3/D7 case we see there is a long tail near zero T, the end
point of which corresponds to the BKT transition. However even
infinitesimal temperature turns it into mean-field type second
order transition\cite{Karch2,Liu}. In Fig \ref{Fig3}bc we plot the
condensate against $\mu$ at a very low temperature ($\tw_H =
10^{-5}$)to show the second order nature. \bigskip \bigskip

{\bf Acknowledgements:} NE and KK are grateful for the support of
an STFC rolling grant.  
KK would like to thank Kristan Jensen and Veselin Filev for discussions.
AG and MM are grateful for University of
Southampton Mayflower Scholarships.

\end{document}